\documentstyle[aps,prabib,psfig,epsf,axodraw]{revtex}
\onecolumn 
\tighten
\begin{document}

\title{
\begin{flushright}
\vspace{-1cm}
{\normalsize MC/TH 98/12 \\
  hep-ph/\
}
\vspace{1cm}
\end{flushright}
Linear Sigma Model and Chiral Symmetry \\
at Finite Temperature 
}

\author{Nicholas Petropoulos}

\address{ Theoretical Physics Group, Department of Physics and Astronomy,\\
University of Manchester, Manchester M13 9PL, UK }

\date{July 9, 1998}

\maketitle
\newcommand{\bold}[1]{\mbox{\boldmath $#1$}}    
\newcommand{\q}{{\bf q}}
\newcommand{\k}{{\bf \rm k}}
\newcommand{\Det}{\rm det}
\newcommand{\beq}{\begin{equation}} 
\newcommand{\eeq}{\end{equation}}
\newcommand{\beqar}{\begin{eqnarray}} 
\newcommand{\eeqar}{\end{eqnarray}}
\newcommand{\Tr}{\rm Tr}
\newcommand{\MeV}{\rm MeV}

\begin{abstract}
The chiral phase transition
is  investigated within the framework of 
the linear sigma model at finite temperature. We  concentrate on
the meson sector of the model and calculate the finite temperature 
effective potential in the Hartree approximation 
by using the Cornwall-Jackiw-Tomboulis formalism of composite
operators. The effective potential is calculated for $N=4$ involving 
the usual sigma and three pions and in the large 
$N$ approximation involving $N-1$ pion fields. In the $N=4$ case we have
examined the theory both in the chiral limit and with the presence
of a symmetry breaking term which generates the pion masses. In both cases 
we have solved the system of the resulting  gap equations
for the thermal effective masses of the particles numerically and we have 
investigated the evolution of the effective 
potential. In the $N=4$ case there is indication
of a first order phase transition and the Goldstone
theorem is not satisfied. The situation is
different in the general case using the large $N$ approximation, the Goldstone 
theorem is satisfied and the phase transition  
is of the second order. For this analysis we have ignored quantum
effects and we used the imaginary time formalism for calculations.
\end{abstract}
 
\section{Introduction}

The study of matter at very high  temperatures and densities 
and of the phase transitions which take place between 
the different phases is very interesting 
from several points of view and it has been the subject
of intense study the latest years since it is relevant to  particle 
physics, astrophysics and cosmology.
According the standard
big bang model it is believed that a series of phases 
transitions happened at 
the early stages of the evolution of the universe, the QCD phase
transition being one of them 
\cite{GPY,shuryak,smilga,linderep,lindebook,kolb}.
There is 
present hope that it could also be
possible to probe the underlying physics of
QCD in the laboratory in experiments involving
relativistic heavy ions collisions. These experiments
are planned for the near future 
and the results could be possible to shed some 
light in questions like the restoration or not of 
chiral symmetry, the nature of quark gluon plasma
and the physics of neutron stars \cite{rajagopal}.

Our aim is to study the chiral symmetry of QCD which is spontaneously 
broken by the small current quark masses. A powerful method in 
approaching questions like 
the restoration of spontaneously broken 
symmetries is to construct order parameters which
characterise the state of symmetry of the system
under consideration. These quantities are zero 
in the one phase but not in the other. A
classic example of an order parameter is the
magnetisation of a ferromagnetic substance, it
is non-zero below  the Curie temperature, but it
disappears at temperatures higher than that. The system
undergoes a transition from an asymmetric,  ordered 
state with non zero magnetisation at low temperature  to  a 
symmetric disordered  state with zero magnetisation
at high temperatures
well above the Curie point. We usually encounter two types of 
phase transitions. In transitions of the first order the
order parameter ``jumps'' discontinuously from its value in the 
one phase to that in the other (usually zero). In contrast 
during second order transitions the order parameter 
vanishes continously.

An important order parameter which is related to the chiral phase
transition of QCD is the 
quark condensate, a measure of the density of quark-antiquark pairs 
that have condensed into the same
quantum mechanical state. They fill the lowest energy state
-the vacuum of QCD- 
and as a result the chiral symmetry 
is broken, since there is no invariance under
chiral transformations \cite{mike}.
We expect
that if we raise the temperature the quark
condensate will disappear and the theory
will be chirally symmetric.

As in many other cases in physics, in order to deal with 
the chiral phase transition we can use effective models to describe
the physical situation. In the case of chiral symmetry
a model with the correct chiral properties is 
the {\it linear sigma model}, a theory of
fermions (quarks or nucleons) interacting with
mesons \cite{Gell-Mann}. This model has been used extensively 
as an effective theory
in low energy phenomenology of QCD describing the physics of 
mesons and it is well suited for
a study of the chiral phase transition \cite{rajagopal,mike}. 
We review briefly the
meson sector of the model and make contact with the 
pion phenomenology  in section 3.

In studies of phase transitions the finite temperature 
effective potential is an
important and popular  theoretical tool. Early stages of applying
similar techniques
goes back at seventies when 
Kirzhnits and Linde \cite{kili} 
were the first 
proposed that  symmetries broken at zero temperature 
could be restored at finite (high enough) 
temperatures. Subsequent work 
by Weinberg \cite{wei}, Dolan and Jackiw \cite{doja}
as well as many others resulted a wide adoption 
of the effective potential 
as the basic tool in such studies. The  
finite temperature effective potential 
$V(\phi,T)$  is defined through an effective 
action $\Gamma (\phi)$  which is the 
generating functional  of the one particle irreducible 
graphs and it has the meaning of 
the free energy density of the system under consideration.

A generalised version is the  
effective potential $V(\phi,G)$ for
composite operators introduced by Cornwall, Jackiw and 
Tomboulis (CJT) \cite{cjt}. 
According to their formalism a generalised version   
of the effective action is introduced, which in contrary to the usual 
effective action $\Gamma (\phi)$  depends not only on 
$\phi(x)$ but  on $G(x,y)$ as well. These two quantities are to be realized 
as the possible expectation values of a quantum field  $\phi(x)$ and the
time ordered product of the field operator $T\phi(x)\phi(y)$ 
respectively. In this case
the effective action $\Gamma(\phi,G)$ is the generating 
functional of the two particle
irreducible vacuum graphs (a graph is called ``two particle 
irreducible'' if it does not become disconnected upon opening two
lines \cite{cjt}). This formalism was originally written at zero
temperature but it has been extended at finite temperature 
by Amelino-Camelia and Pi 
where it was used 
for investigations of the effective potential of the 
$\lambda\phi^4$ theory \cite{gacpi} and gauge 
theories \cite{gac94}.

Physical solutions demand minimization of the effective 
action  with respect to both $\phi$ and $G$ \cite{cjt,gacpi}. As a result 
the  CJT effective potential should
satisfy the stationarity requirments
\begin{equation}
\frac{d V(\phi,G)}{d\phi}=0
\end{equation}
and
\begin{equation}
\frac{dV(\phi,G)}{d G}=0~.
\end{equation}  
Then the  conventional effective potential results 
as $V(\phi)= V(\phi;G_{0}(\phi))$ at the solution 
$G(\phi)=G_{0}(\phi)$ of the second equation.

There is an advantage in using the CJT method to calculate the 
effective potential in  Hartree approximation.  According to reference
\cite{gacpi}, using an ansantz for a  ``dressed propagator''
we need to evaluate only one graph that of the ``double bubble''
in fig. 1a, instead of summing  the infinite class of ``daisy'' and 
``super daisy'' graphs, given in figs. 1b, 1c  respectively, using  the usual 
tree level propagators.
\begin{figure}
\begin{center}
\begin{picture}(300,100)(0,0)
\BCirc(35,40){10}
\BCirc(35,60){10}
\Text(35,10)[c]{(a)}
\Text(150,10)[c]{(b)}
\Text(260,10)[c]{(c)}
\BCirc(150,50){20}
\CArc(167.7,67.7)(5,0,360)
\CArc(132.3,67.7)(5,0,360)
\CArc(167.7,32.3)(5,0,360)
\CArc(132.3,32.3)(5,0,360)
\BCirc(150,75){5}
\BCirc(150,25){5}
\BCirc(175,50){5}
\BCirc(125,50){5}
\CArc(167.7,67.7)(5,0,360)
\CArc(167.7,67.7)(5,0,360)
\CArc(167.7,67.7)(5,0,360)
\BCirc(260,50){20}
\BCirc(290,50){10}
\BCirc(305,50){5}
\BCirc(312.5,50){2.5}
\BCirc(290,65){5}
\BCirc(290,72.5){2.5}
\BCirc(260,80){10}
\BCirc(260,95){5}
\BCirc(260,102.5){2.5}
\BCirc(275,80){5}
\BCirc(245,80){5}
\end{picture}
\end{center}
\caption{The double bubble and examples of daisy and superdaisy diagrams}
\end{figure}
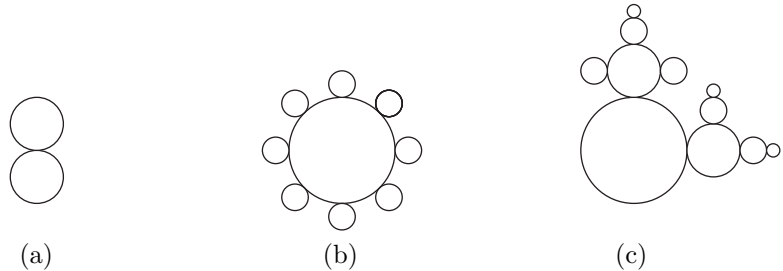

We demonstrate
the advantage of this method in the next section  where we
calculate the finite temperature effective potential for one 
scalar field with quartic self-interaction. The calculation of
 the effective potential 
by using the CJT formalism is reviewed in great detail in
references \cite{gacpi,gac96} 
but in order to illustrate
the method for the calculation of the effective 
potential for the linear sigma model at finite temperature 
we will reproduce here the basic steps.

In our calculations we use the imaginary time formalism, known as
Matsubara formalism \cite{doja,kapusta,lebellac,landsman}. According to 
this technique 
we work in Euclidean space-time and
use the  same Feynman rules as at zero temperature
but evaluating momentum space integrals we replace integration
over the time component $k_{4}$ with a summation over discrete
frequencies which means that  in the case of 
bosons  $k_{4}=2\pi i n T,\;\;n=0,\pm 1,\pm 2,\pm 3,\ldots$.
This is encoded into the following relationship
\begin{equation}
\int\frac{d^{4}k}{(2\pi)^{4}} f(k)\longrightarrow
 \frac{1}{\beta}\sum_{n}\int\frac{d^{3}{\bf k}}{(2\pi)^{3}} 
f(2\pi i n T,{\bf k})
\equiv\int_{\beta}f(2\pi i n T,{\bf k})~,
\end{equation}
where $\beta$ is the inverse temperature, $\beta=1/k_{B}T$, and
as usual Boltzmann's constant is taken $k_{B}=1$. For the shake of 
simplicity in 
what follows we have introduced a subscript
$\beta$ to denote integration and summation over the Matsubara
frequency sums.

The outline of the remaining 
sections is as follows. In next section 2 we calculate the effective 
potential for the $\lambda \phi^{4}$ theory using the CJT
method. In section 3 we apply the same technique
to calculate the effective potential for the linear sigma model
in the Hartree approximation and we solve numerically the system 
of the resultant
gap equations both in the chiral limit and with the presence of
a linear term which breaks the chiral symmetry 
of the Lagrangian and generates the 
pion observed masses. We have repeated these steps for the 
generalised version of the linear sigma model with $N-1$ pion fields  
in the large $N$ approximation. Finally, in 
section 4 we outline 
our results and conclusions.

\section{The $\lambda\Phi^{4}$ theory}

The Euclidean Lagrange 
density for a single scalar field with quartic self interaction
interaction is given by
\begin{equation}
{\cal L} = {1 \over 2} (\partial_{\mu} \Phi) (\partial^{\mu} \Phi)
-{1 \over 2} m^2 \Phi^2
-{\lambda \over 24}  \Phi^4 ~,
\label{lf}
\end{equation}
where the ``mass squared'' $m^2$ is considered as a negative parameter, in
order to realise the spontaneous breaking of symmetry.
By shifting the field as $\Phi\rightarrow \Phi+\phi$ the 
``classical potential'' takes the form 
\begin{equation}
U(\phi)={1 \over 2} m^2 \phi^2
+{\lambda \over 24}  \phi^4~,
\label{classical}
\end{equation}
and the interaction Lagrangian which 
describes the vertices of the shifted 
theory is given by
\begin{equation}
{\cal L}_{int}(\sigma;\phi) =
-{\lambda \over 6} \phi\Phi^3 - {\lambda \over 24} \Phi^4 ~.
\end{equation}
The tree-level propagator which corresponds to the 
above Lagrangian density is
\begin{equation}
{\cal D}^{-1}(\phi;k) = k^2 +m^2 + {1\over 2}\lambda\phi^2  ~.
\end{equation}

According to CJT formalism \cite{gacpi} 
the finite temperature effective potential is given by 
\begin{eqnarray}  
V(\phi,G) &=& U(\phi)+{1 \over 2}\int_{\beta}\ln 
              G^{-1}(\phi;k)\nonumber\\
          & &+{1 \over 2}\int_{\beta}
              [{\cal D}^{-1}(\phi;k)G(\phi;k)-1]\nonumber\\
          & &+V_2(\phi,G)~,
\end{eqnarray}
where $U(\phi)$ is the ``classical potential'' given by
equation (\ref{classical})
and $V_2(\phi,G)$ represents the infinite sum of the 
two-particle irreducible vacuum graphs. We are 
going to evaluate the effective potential
in the Hartree approximation which corresponds to that 
the leading contribution to last term $V_{2}(\phi,G)$ of
the effective potential comes from the ``double bubble''
diagram given in fig. 1a \cite{cjt,gacpi}. Therefore the effective 
potential results as
\begin{eqnarray}
V(\phi,G) &=& {1 \over 2}m^2 \phi^2 
+ {1 \over 24}\lambda \phi^4
+{1 \over 2} \int_{\beta} \ln G^{-1}(\phi;k) 
\nonumber\\
& & +{1 \over 2}\int_{\beta} 
[(k^2 + m^2 + {1\over 2}\lambda  \phi^2)G(\phi;k) -1] 
+ {1 \over 8}\lambda 
\left[ \int_{\beta}  G(\phi;k) \right]^2 ~. 
\label{potential}
\end{eqnarray}
Minimizing the effective potential with respect
to ``dressed propagator'' $G(\phi;k)$ we obtain a gap equation
\begin{equation}
G^{-1}( \phi;k) = k^2 + m^2 + {1\over 2}\lambda \phi^2
+ {1 \over 2}\lambda\int_{\beta} G(\phi;k) ~. 
\end{equation}
The solution $G_0(\phi;k)$  of the gap equation is inserted back
into the expression for the effective potential to give the potential
as a function of $\phi$. According to \cite{cjt} using 
the propagator $G_{0}(\phi;k)$ for internal lines 
corresponds to summing all daisy and super-daisy
diagrams using the usual tree level propagators as in \cite{doja}.

In order to proceed 
and according to reference \cite{gacpi}, it is convenient to 
adopt the following form for the
propagator $G(\phi;k)$
\begin{equation}
G(\phi;k)= {1 \over k^2 + M^2 }~,
\end{equation}
where an ``effective mass'' $M=M(\phi;k)$ has been 
introduced. Then 
the gap equation for the propagator becomes an equation
for the effective mass
\begin{equation}
M^2 = m^2+{\lambda \over 2}\phi^2+ 
{\lambda \over 2}\int_{\beta}{1 \over k^2 + M^2 }~,
\label{gap}
\end{equation}
where it is obvious that in this approximation
the effective mass $M$ is momentum independent.

In terms of the solution $M_{0}(\phi)$ of  the gap 
equation (\ref{gap}), the
effective potential takes the form
\begin{eqnarray}
V( \phi,M_{0}) &=& {1 \over 2} m^2 \phi^2
+{\lambda \over 24}\phi^4+{1 \over 2}\int_{\beta}\ln 
              (k^{2}+M^{2}_{0})\nonumber\\
          & &-{1 \over 2}(M^2_{0} - m^2 - {\lambda \over 2} \phi^2 )
\int_{\beta}{1 \over k^2 + M^2_{0} }
             + {1 \over 8}\lambda\left[ \int_{\beta}
              {1 \over k^2 + M_{0}^2}\right ]^{2}~. 
\end{eqnarray}

Performing the Matsubara frequency sums as in \cite{doja}, the 
logarithmic integral which appear into 
the above expression of the effective potential
divides into a zero temperature part $Q_{0}(M)$ which is divergent 
and one non zero $Q_{\beta}(M)$
temperature part which is finite and can be written as
\begin{eqnarray}
Q(M)&=&{1\over 2}\int_{\beta}\ln(k^{2}+M^{2})=Q_{0}(M)+
Q_{\beta}(M)\nonumber\\
&=&\int\frac{d^{3}{\bf k}}{(2\pi)^{3}}\frac{\omega_{\rm k}}{2}+\frac{1}{\beta}
\int\frac{d^{3}{\bf k}}{(2\pi)^{3}}
\ln[1-\exp(-\beta\omega_{\rm k})]~,
\end{eqnarray}
where $\omega_{\rm k}=({\bf k}^{2}+M^{2})^{1/2}$. In this above 
expression and in what follows we omit the subscript 0 of $M$.
Similarly, the second integral is divided into two parts as well. One zero 
temperature part $F_{0}(M)$ and a finite temperature  $F_{\beta}(M)$  part as
\begin{eqnarray}
F(M)&=&\int_{\beta} {1 \over k^2 + M^2 }=
F_{0}(M)+F_{\beta}(M)\nonumber\\
&=&\int\frac{d^{3} {\bf k}}{(2\pi)^{3}}\frac{1}{2\omega_{\rm k}}
+\frac{1}{\beta}
\int\frac{d^{3}{\bf k}}{(2\pi)^{3}}\frac{1}{\omega_{\rm k}}
\frac{1}{\exp(-\beta\omega_{\rm k})-1}~.
\end{eqnarray}
The second term vanishes at zero temperature, while 
the first term survives but it gives
rise to divergences which can be carried out 
using appropriate renormalisation prescriptions \cite{gacpi,gac96}.
If one is interested  in temperature induced  
effects only, as it is our
approximation, the divergent integrals can be ignored. In this case, by 
making a change of the
integration variables
the finite temperature part of  $F(M)$ can be written as
\begin{equation}
F_{\beta}(M)=\frac{T^{2}}{2\pi^{2}}\int_{0}^{\infty}\frac{x^2 dx}
{[x^{2}+y^{2}]^{1/2}}
\frac{1}{\exp[x^{2}+y^{2}]^{1/2}-1}~,
\label{FM}
\end{equation}
where we have used a shorthand notation and $y=M/T$. Similarly 
the finite temperature part of the logarithmic integral
becomes
\begin{equation}
Q_{\beta}(M)=\frac{T^4}{2\pi^2}\int_{0}^{\infty}dx x^2 
\ln\left [1-\exp(-[x^2+y^2]^{1/2})\right ]~.
\label{QM}
\end{equation} 
Then the finite 
temperature effective potential, ignoring quantum 
effects, can be written as 
\begin{eqnarray}
V(\phi,M)&=&\frac{1}{2}m^{2}\phi^{2}+\frac{1}{24}\lambda\phi^{4}
+ Q_{\beta}(M)\nonumber\\
&&\mbox{}-\frac{1}{2}(M^{2}-m^{2}-\frac{\lambda}{2}\phi^{2})F_{\beta}(M)
+\frac{\lambda}{8}[F_{\beta}(M)]^{2}~.
\end{eqnarray}
We can obtain a more compact form if we make use of the gap equation 
(\ref{gap}), 
\begin{equation}
V(\phi,M)=\frac{1}{2}m^{2} \phi^{2}+\frac{1}{24}\lambda \phi^{4}
+ Q_{\beta}(M)-\frac{1}{8}\lambda\left[F_{\beta}(M)\right]^{2}~.
\end{equation}

\section{The Effective Potential of the Linear Sigma Model}

\subsection{The Linear Sigma Model}

As we mentioned in the introduction, the linear sigma model
serves as a good low energy effective theory 
in order one to have some  insight into QCD.
The model was first introduced in the sixties \cite{Gell-Mann} 
as a model for pion-nucleon interactions and has attracted much 
attention recently
specially in studies involving the Disoriented Chiral
Condensates (DCC's) \cite{bjorken,mikeaba,randrup,randrup-nucl,rw404}.
The model is very
well suited for 
describing the physics of pions in studies
of chiral symmetry. Fermions are inserted in the model either
as nucleons, if one is to study nucleon 
interactions or as quarks.  The mesonic part of the model 
consists of four scalar fields,  one
scalar isoscalar field which is called the  sigma $\sigma$ field  and 
the usual three pion fields $\pi^{0}$, $\pi^{\pm}$ which form a 
pseudo-scalar isovector. The fields form a 
four vector $(\sigma,\pi_{i}),\;\;i=1,2,3$ which we regard as the
chiral field and the model
displays an $O(4)$ symmetry. The $\sigma$ field can be used to represent
the quark condensate, the order parameter for
the chiral phase transition,  since they exhibit the same behaviour
under chiral transformations \cite{mike}. The pions 
are very light particles and can
be considered  approximately as massless Goldstone 
bosons. We use the model as an effective
theory for QCD ignoring
the fermion sector at the moment and 
concentrating on the meson sector.

The generalised version 
of the meson sector 
of the linear sigma model called the $O(N)$
or vector model and is based on a set of $N$ 
real scalar fields. The $O(N)$ model Lagrangian 
can be written as
\begin{equation}
\label{nmodel}
{\cal L}=\frac{1}{2}(\partial_{\mu}{\bf \Phi})^{2}-
\frac{1}{2}m^{2}{\bf \Phi}^{2}
-\frac{1}{6N}\lambda {\bf \Phi}^{4}-\varepsilon\sigma~,
\end{equation}
and in the absence of the last term it remains invariant 
under  $O(N)$ symmetry 
transformations
for any $N\times N$ orthogonal matrix.
In order our notation to be consistent with applications on pion 
phenomenology we can identify the $\Phi_{1}$ with the $\sigma$
field and the remaining $N-1$ components as the pion 
fields, that is ${\bf \Phi}=(\sigma, \pi_{1},\ldots,\pi_{N-1})$. The 
last term, $\varepsilon \sigma$ into the above expression 
has been introduced in order to generate the  observed masses
of the pions. 

\noindent
The contact with phenomenology is obtained by considering the
the case $N=4$. Then the  Lagrangian of the model is given by
\begin{equation}
{\cal L}=\frac{1}{2}(\partial\sigma)^2 + \frac{1}{2}(\partial \bold {\pi})^2
        -\frac {1}{2}m^{2}\sigma^2 -\frac {1}{2}m^{2}
 \bold {\pi}^2 - \frac{\lambda}{24}(\sigma^2 +
 \bold {\pi}^2)^2 - \varepsilon \sigma~,
\label{lagrangian4}
\end{equation} 
where $\varepsilon=f_{\pi}m_{\pi}^{2}$
and $f_{\pi}=93\,\,\rm MeV$ is the pion decay constant. At zero temperature
and in order to be consistent with the pion observed mass of
$m_{\pi}\approx 138\;\;\rm MeV$ and the usually adopted sigma mass
$m_{\sigma}\approx 600\;\;\rm MeV$ we choose the coupling 
constant $\lambda$ in our model to be
\begin{equation}
\lambda=\frac{3(m_{\sigma}^{2}-m_{\pi}^{2})}{f_{\pi}^{2}}~.
\end{equation}
The negative mass parameter $m^{2}$ has been introduced in order
to obtain spontaneous breaking of symmetry and its value is chosen to be
\begin{equation}
-m^{2}=(m_{\sigma}^{2}-3 m_{\pi}^{2})/2 > 0~.
\end{equation}
As we have referred earlier, our aim is to use the linear
sigma model as a model to study the chiral phase transition, so
we need to calculate the effective potential. For this 
calculation we adopt the CJT 
formalism which has been presented earlier for the $\lambda\phi^{4}$
theory and calculate the effective potential in the Hartree and
large $N$ approximations.

\subsection{Hartree approximation in the chiral limit $\varepsilon=0$}

In order to deal with the exact chiral limit first, the starting point 
is the Lagrangian (\ref{lagrangian4}) and we  ignore
the symmetry breaking term at the moment.
By shifting the sigma field as $\sigma\rightarrow \sigma+\phi$ 
the ``classical potential'' results as
\begin{equation}
U(\phi)=\frac{1}{2}m^{2}\phi^{2}+\frac{\lambda}{24}\phi^{4}~,
\end{equation}
and the interaction 
Lagrangian which 
describes the vertices of the new 
theory takes the form
\begin{equation}
{\cal L}_{int}(\sigma;\phi) =
- {\lambda \over 4!} \sigma^4 
-{\lambda \over 4!}{\bf \pi}^4
-{\lambda \over 12}\sigma^2{\bf \pi}^2
-{\lambda \over 6} \phi\sigma^3 
-{\lambda \over 6} \phi\sigma{\bf\pi}^{2} ~,
\label{lint}
\end{equation}
plus terms linear in the $\sigma$ field and constants which we omit
for simplicity. In our approximation we do not consider interactions
given by the last two terms in the Lagrangian.

The tree level sigma and pion propagators 
corresponding to the above Lagrangian are
\begin{equation}
{\cal D}^{-1}_{\sigma}(\phi;k) = k^2 +m^2 + {1\over 2}\lambda\phi^2  ~,
\label{Ds}
\end{equation}
\begin{equation}
{\cal D}^{-1}_{\pi}(\phi;k) = k^2 +m^2 + {1\over 6}\lambda\phi^2  ~.
\label{Dp}
\end{equation}
We evaluate the effective potential in the Hartree
approximation which means that we only need  to calculate 
the ``double bubble''
diagrams as in $\lambda\phi^{4}$ theory. In the linear sigma 
model the corresponding 
effective potential at finite temperature  can be written as
\begin{eqnarray}
V(\phi,G) &=& U(\phi)+{1 \over 2}\int_{\beta}\ln 
              G^{-1}_{\sigma}(\phi;k)
+{3 \over 2}\int_{\beta}\ln G^{-1}_{\pi}(\phi;k)
\nonumber\\
          & &+{1 \over 2}\int_{\beta}
              [{\cal D}_{\sigma}^{-1}(\phi;k)G_{\sigma}(k)-1]
+{3 \over 2}\int_{\beta}
              [{\cal D}_{\pi}^{-1}(\phi;k)G_{\pi}(\phi;k)-3]\nonumber\\
          & &+V_2(\phi,G_{\sigma},G_{\pi})~,
\end{eqnarray}
where the first term $U(\phi)$ is the ``classical potential''
and the last term $V_2(\phi,G_{\sigma},G_{\pi})$ originates from the
sum of ``double bubble'' diagrams \cite{gacpi}. There are four types
of double bubbles as we show in fig. 2 and these contribute 
the following terms in the potential
\begin{equation}
V_2(\phi,G_{\sigma},G_{\pi})=3\frac{\lambda}{24}\left [
\int_{\beta}G_{\sigma}(\phi;k)\right ]^{2}
+15\frac{\lambda}{24}\left [
\int_{\beta}G_{\pi}(\phi;k)\right ]^{2}
+6\frac{\lambda}{24}\left [\int_{\beta}G_{\sigma}(\phi;k)\right ]
\left [\int_{\beta}G_{\pi}(\phi;k)\right ]~.
\eeq
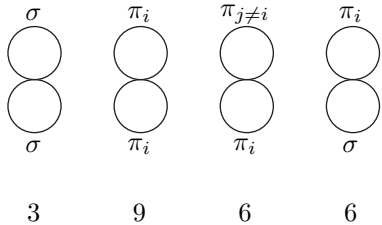
\begin{figure}
\label{bubbles}
\begin{center}
\begin{picture}(300,100)(0,0)
\BCirc(85,55){10}
\BCirc(85,75){10}
\Text(85,40)[c]{$\sigma$}
\Text(85,90)[c]{$\sigma$}
\Text(85,15)[c]{$3$}
\BCirc(125,55){10}
\Text(125,40)[c]{$\pi_{i}$}
\BCirc(125,75){10}
\Text(125,90)[c]{$\pi_{i}$}
\Text(125,15)[c]{$9$}
\BCirc(165,55){10}
\Text(165,40)[c]{$\pi_{i}$}
\BCirc(165,75){10}
\Text(165,90)[c]{$\pi_{j\neq i}$}
\Text(165,15)[c]{$6$}
\BCirc(205,55){10}
\Text(205,40)[c]{$\sigma$}
\BCirc(205,75){10}
\Text(205,90)[c]{$\pi_{i}$}
\Text(205,15)[c]{$6$}
\vspace {20pt}
\end{picture}
\caption{The double bubble graphs which contribute to
the effective potential for the O(4) linear sigma model in 
the Hatree approximation. The numbers show the weight
of each type of bubble in the expression for the effective
potential.} 
\end{center}
\end{figure}

Minimizing the effective potential with respect 
to the ``dressed'' propagators,  we get the following set 
of nonlinear gap equations
\begin{eqnarray} 
&&G_{\sigma}^{-1}(\phi;k)={\cal D}_{\sigma}^{-1}(\phi;k)+\frac{\lambda}{2}
\int_{\beta}G_{\pi}(\phi;k)
+\frac{\lambda}{2}\int_{\beta}G_{\sigma}(\phi;k)
\nonumber\\
&&G_{\pi}^{-1}(\phi;k)={\cal D}_{\pi}^{-1}(\phi;k)+\frac{5\lambda}{6}
\int_{\beta}G_{\pi}(\phi;k)
+\frac{\lambda}{6}
\int_{\beta}G_{\sigma}(\phi;k)~.
\label{gap-prop}
\end{eqnarray}

In order to proceed we can use  the same 
ansatz for the dressed propagators as 
in the one field case \cite{gacpi}
\begin{equation}
G_{\sigma/\pi}^{-1}=k^{2}+M^{2}_{\sigma/\pi}~,
\label{dressed} 
\end{equation}
and using the equations (\ref{Ds}), (\ref{Dp}), (\ref{gap-prop}) 
and (\ref{dressed}) we end up
with the following system for the thermal
effective  masses 
\begin{eqnarray}
&&M_{\sigma}^{2}=m^{2}+\frac{1}{2}\lambda\phi^{2}+
\frac{\lambda}{2}F(M_{\sigma})+\frac{\lambda}{2}F(M_{\pi})
\nonumber\\
&&M_{\pi}^{2}=m^{2}+\frac{1}{6}\lambda\phi^{2}+\frac{\lambda}{6}F(M_{\sigma})
+\frac{5\lambda}{6}F(M_{\pi})~.
\label{system}
\end{eqnarray}
In these last two equations we have used a  shorthand
notation and $F(M)$ is given by
\begin{equation}
F(M)=\int_{\beta}\frac{1}{k^{2}+M^{2}}~.
\end{equation}
As in $\lambda\phi^{4}$ theory the thermal effective masses  are 
independent of momentum and functions of the order
parameter $\phi$ and the 
temperature $T$.

By using these two equations
the effective potential at finite  temperature 
then can be written as
\begin{eqnarray}
V(\phi,M)&=&\frac{1}{2}m^{2}\phi^{2}+\frac{1}{24}\lambda\phi^{4}
+ \frac{1}{2}\int_{\beta}\ln(k^{2}+M_{\sigma}^{2})
+ \frac{3}{2}\int_{\beta}\ln(k^{2}+M_{\pi}^{2})
\nonumber\\
         &&\mbox{}- \frac{1}{2}(M_{\sigma}^{2}-m^{2}
           -{1\over 2}\lambda\phi^{2})F(M_{\sigma})
   -\frac{3}{2}(M_{\pi}^{2}-m^{2}-\frac{1}{6}\lambda\phi^{2})F(M_{\pi})
\nonumber\\
 &&\mbox{}+\frac{\lambda}{8}[F(M_{\sigma})]^{2}+\frac{5\lambda}{8}
[F(M_{\pi})]^{2}+\frac{\lambda}{4}F(M_{\sigma})F(M_{\pi})~.
\end{eqnarray}
Minimizing the effective potential with respect 
to the ``dressed'' propagators  we  have found  
the set of nonlinear gap equations for the effective particles' masses
given by equation (\ref{system}). In addition, by minimizing 
the potential with respect to the order parameter
 we obtain one more equation
\begin{equation}
0=m^{2}+\frac{1}{6}\lambda\phi^{2}+\frac{\lambda}{2}F(M_{\sigma})
+\frac{\lambda}{2}F(M_{\pi})~.
\label{phi4}
\end{equation}
In order to study the evolution of the potential as a function of 
temperature, we
perform the Matsubara frequency sums \cite{doja} as in 
the one field case. There are some troubles
concerning the renormalisation of the model 
\cite{gacpi,gac96,baym,matsui,gac97}. At the level of our
approximation we ignore 
quantum effects for the moment and keep  only the finite 
temperature part of the 
integrals. Using a compact notation the finite 
temperature effective potential can be written in the  form
\begin{eqnarray}
V(\phi,M)&=&\frac{1}{2}m^{2}\phi^{2}+\frac{1}{24}\lambda\phi^{4}
+ Q_{\beta}(M)
+ 3Q_{\beta}(M_{\pi})\nonumber\\
         &&\mbox{}-\frac{\lambda}{8}[F_{\beta}(M_{\sigma})]^{2}
-\frac{5\lambda}{8}[F_{\beta}(M_{\pi})]^{2}
-\frac{\lambda}{4}F_{\beta}(M_{\sigma})F_{\beta}(M_{\pi})~,
\end{eqnarray}
where in this last step we have used the 
gap equations (\ref{system}). The expessions for 
$F_{\beta}(M)$and $Q_{\beta}(M)$ are
given by the equations (\ref{FM}) and (\ref{QM}).

In order to calculate the effective masses as functions of temperature
we need to solve the system of  equations (\ref{system}) and 
(\ref{phi4}). We first observe that if $\phi=0$, which happens 
in the high temperature
phase, the two equations become degenerate, the particles have 
the same mass and we have to solve only one  equation 
\begin{equation}
M^{2}=m^{2}+\lambda F_{\beta}(M)~,
\end{equation}
where obviously as in the expression of the effective potential we
keep only the finite temperature part of the integral. This last 
equation can be used to define a ``transition temperature'' 
$T_{c1}$. This temperature is defined as the 
temperature where both particles become massless.  Recall now 
that $F_{\beta}(M)$ is given by
\begin{equation}
\label{FBM}
F_{\beta}(M)=\frac{T^{2}}{2\pi^{2}}\int_{0}^{\infty}\frac{x^2 dx}
{[x^{2}+(M/T)^{2}]^{1/2}}
\frac{1}{\exp[x^{2}+(M/T)^{2}]^{1/2}-1}~,
\end{equation}
where we show explicitly the dependence of $F_{\beta}(M)$  
to mass and temperature. When the mass 
of the particles vanishes the integral  
reduces to the well known
\begin{equation}
I(x)=\int_{0}^{\infty}\frac{x dx}{e^{x}-1}=\frac{\pi^{2}}{6}~,
\label{pi/6}
\end{equation}
and the temperature $T_{c1}$ is found to be 
\begin{equation}
T_{c1}=\sqrt{2}\left (-\frac{ 6 m^{2}}{\lambda}\right )^{1/2}~.
\end{equation}
But in  defining our model parameters we have chosen that at zero
temperature 
$\phi^{2}=f_{\pi}^{2}= -6 m^{2}/\lambda$, where 
$f_{\pi}=93\; \rm MeV$  is the 
pion decay constant, so we find that
$T_{c1}=\sqrt{2} f_{\pi} \approx 131.5\; \rm MeV$. 

In the low temperature phase we can
eliminate  $\phi$ and end up 
with the following nonlinear system 
\begin{eqnarray}
&&M_{\sigma}^{2}=-2m^{2}-\lambda F_{\beta}(M_{\sigma})
-\lambda F_{\beta}(M_{\pi})
\nonumber\\
&&M_{\pi}^{2}=-\frac{\lambda}{3}F_{\beta}(M_{\sigma})
+\frac{\lambda}{3}F_{\beta}(M_{\pi})~,
\label{system1}
\end{eqnarray}
which we solve numerically and the solution is presented in fig. 3a.
\begin{figure}[h]
\begin{center}
\mbox{
\epsfxsize=10cm
\epsfbox{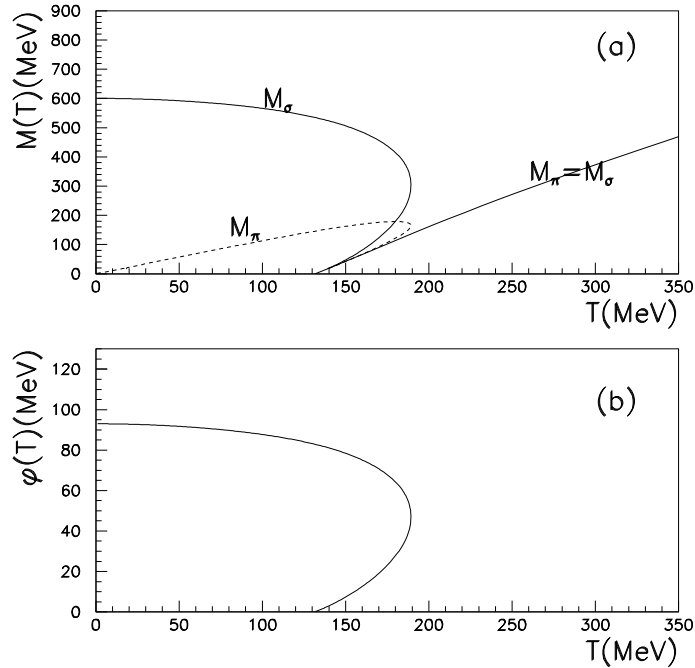}
}
\end{center}
\caption{(a) Solution of the system of gap equations in the chiral limit. The
sigma and pion effective masses are given as functions
 of temperature. (b) Evolution of the order parameter 
as a function of temperature.}
\end{figure}

\noindent
As shown in fig. 3a, the temperature $T_{c1}$ which 
is calculated numerically was found to be  in excellent agreement with 
the value  obtained by using the limit of 
the  high temperature equations with degenerate masses. 

At this point we can observe that 
there is an indication of a first order phase transition, because 
combining the first of the two equations in (\ref{system}) with
equation (\ref{phi4}) we find that
\begin{equation}
M_{\sigma}^{2}=\frac{1}{3}\lambda\phi^{2}~.
\end{equation}
This last equation shows of course that the 
order parameter varies with temperature proportionally
to the sigma mass. The temperature dependence of $\phi=\phi(T)$
is calculated by using the sigma mass as it was found
by solving the system in eqn (\ref{system1}). This is given in fig. 3b
where it is obvious that this approximation predicts
a first order phase transition. This last observation coincides with 
the qualitative picture given
by Baym and Grinstein in their early paper \cite{baym} where their
``modified Hartree approximation'' predicts a first order phase
transition as well. However, in contrast to our 
approximation into their analysis they included quantum 
effects as well. Signals of a first order phase transition have also
been reported  in recent analyses by Randrup \cite{randrup-linear}
and by Roh and Matsui \cite{matsui}.

In order to get more  insight into the nature of the phase transition 
and verify that the transition is of the first order
we can calculate the effective potential $V(\phi,T)$ as a function
of the temperature and the order parameter. So, we first solve 
numerically  the system of equations
(\ref{system}) and (\ref{phi}) (where of course we keep only
the finite temrerature part of the integrals) and calculate the 
effective masses 
of the particles as functions the order parameter and the temperature. Finally
the effective potential is calculated numerically by using
these masses. The evolution of the potential for several temperatures
is given in  fig. 4. The
shape of the potential confirms that 
a first order phase transition takes place, since it exhibits two 
degenerate minima at a 
temperature $T_{c}\approx 182\;{\rm MeV}$
which is usually defined as the transition 
temperature. The second minimum of 
the potential at $\phi \neq 0$ disappears at a temperature  
$T_{c2}\approx 187\;{\rm MeV}$. The temperatures $T_{c1}$ 
and $T_{c2}$ are called (in condensed matter terminology) 
the lower and upper spinodal 
points respectively. Between these temperatures metastable 
states exist and the system can exhibit supercooling or superheating. For
$T_{c1}< T < T_{c}$ the metastable states are centered around the 
origin since for $T_{c}< T < T_{c2}$ the metastable states occur
for $\phi \neq 0$. When the system reaches $T_{c1}$ or $T_{c2}$ 
the curvature of the
potential at the metastable minima vanishes. A discussion about 
first order phase 
transitions and more details about how these
transitions procceed can be found in refs \cite{linderep,lindebook,kolb}.  
\begin{figure}[h]
\label{veff}
\begin{center}
\mbox{
\epsfxsize=10cm
\epsfbox{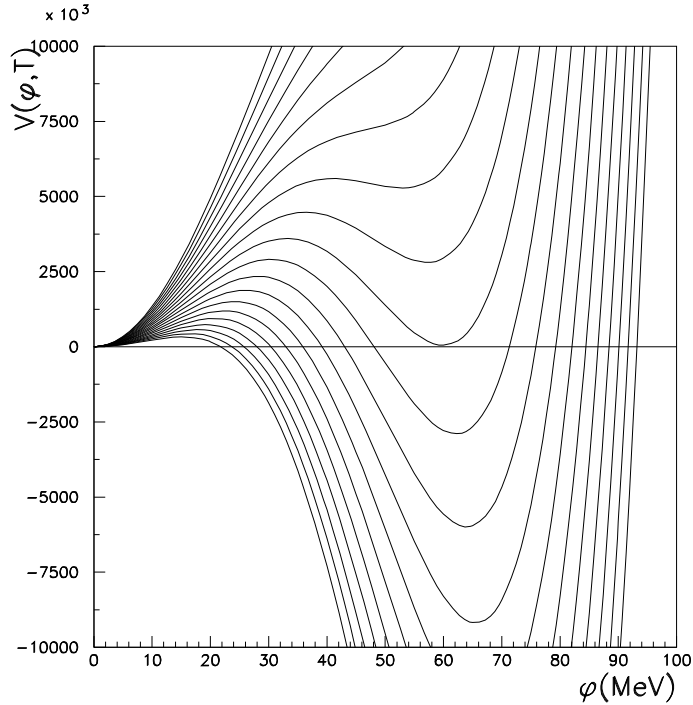}
}
\end{center}
\caption{Evolution of the effective potential $V(\phi,T)$ as a function of 
the order parameter $\phi$ for several temperatures
in steps of $2\;{\rm MeV}$. The two minima appear 
as degenerate at $T_{c}\approx 182\; {\rm MeV}$.} 
\end{figure}

\subsection{Hartree approximation in the broken symmetry 
case $\varepsilon \neq 0$}

When $\varepsilon \neq 0$
the term linear in the sigma field 
into the Lagrangian 
generates the pion observed masses. This term 
is $G$ independent and so
minimization of the potential with respect to ``dressed'' propagators
will give us the same set of gap equations for the effective
masses as before. However, minimizing the potential with respect to $\phi$ we
get the following equation
\begin{equation}
\left [ m^{2}+\frac{1}{6}\lambda\phi^{2}+\frac{\lambda}{2}F_{\beta}(M_{\sigma})
+\frac{\lambda}{2}F_{\beta}(M_{\pi})\right ]\phi-\varepsilon =0~.
\label{phi}
\end{equation}
In order to proceed we need to solve the nonlinear system of  three
equations (\ref{system}) and (\ref{phi}). We first observe that at $T=0$
the above equation becomes
\begin{equation}
M_{\pi}^{2}=m^{2}+\frac{1}{6}\lambda\phi^{2}=\frac{\varepsilon}{\phi}
=m_{\pi}^{2}~,
\end{equation}
where $m_{\pi}$ is the tree level pion mass. Then for
$\phi=f_{\pi}$ we recover 
the relation between the pion mass at zero temperature and the 
symmetry breaking 
factor $\varepsilon$: $\varepsilon=f_{\pi}m_{\pi}^{2}$ 
where $f_{\pi}$ is the pion decay constant.
We solved the system of eqns (\ref{system}) and (\ref{phi})
numerically and the solution is presented 
in fig. 5a. At low temperatures the pions appear with
the observed masses but their mass increases with temperature
since the sigma mass decreases. At high temperatures (higher
than $\sim 300\; \rm MeV$) due to interactions in the thermal 
bath all particles appear to have the 
same effective mass. 

The appearance of the symmetry breaking term into 
the Lagrangian modifies the evolution of the
order parameter $\phi$
as well. As it is obvious in fig. 5b, as the 
temperature increases the order parameter decreases
and at very high temperatures vanishes smoothly. 
But in this case the change is not a phase transition any more. We rather
encounter a smooth crossover from a low temperature phase when the particles
appear with different masses to a high temperature phase where the
thermal contribution to the effective masses makes them degenerate.

\begin{figure}[h]
\begin{center}
\mbox{
\epsfxsize=10cm
\epsfbox{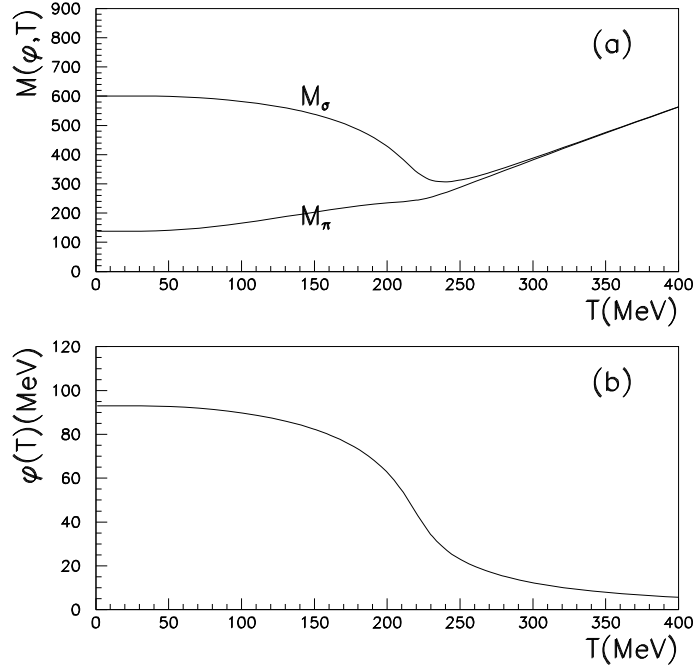}
}
\end{center}
\caption{(a) Solution of the system of gap equations in the case 
when $\epsilon \neq 0$. At low temperatures the pions appear with
the observed masses. (b) Evolution of the order parameter as a function 
of temperature.}
\end{figure}

\subsection{Large $N$ approximation in the chiral limit $\varepsilon=0$}  

The large $N$ approximation of the linear sigma model has been studied
recently by Amelino-Camelia \cite{gac97} and our expressions are very
similar to the ones obtained there, since the same method is used 
in both cases. However, in our approach we do not consider 
the renormalization of the
model because in our approximation we take into account only finite
temperature effects. Our analysis is in a sense complementary to that
in \cite{gac97} since we solve the system of gap equations and
consider the effects  of the symmetry breaking term (the last term in the 
Lagarangian (\ref{nmodel})) which is omitted in reference \cite{gac97}. 

\noindent
The Lagrangian of the linear sigma model when we consider a 
large number ($N-1$) pion
fields is
given in equation (\ref{nmodel}) and  shifting the sigma
field as $\sigma \rightarrow \sigma+\phi$, the tree level 
propagators  are
\begin{equation}
D^{-1}_{\sigma}(\phi;k) = k^2 +m^2 + {2\lambda \over N}\phi^2  ~,
\end{equation}
\begin{equation}
D^{-1}_{\pi}(\phi;k) = k^2 +m^2 + {2\lambda \over 3N}\phi^2  ~.
\end{equation}

Then the  effective potential
at finite temperature will appear as
\begin{eqnarray}
V(\phi,M) &=&\frac{1}{2}m^{2}\phi^{2}+\frac{1}{6N}\lambda\phi^{4} 
+{1 \over 2}\int_{\beta}\ln G^{-1}_{\sigma}(\phi;k)
+{N-1\over 2}\int_{\beta}\ln G^{-1}_{\pi}(\phi;k)
\nonumber\\
          & &+{1 \over 2}\int_{\beta}
              [{\cal D}_{\sigma}^{-1}(\phi;k)G_{\sigma}(\phi;k)-1]
+{N-1 \over 2}\int_{\beta}
              [{\cal D}_{\pi}^{-1}(\phi;k)G_{\pi}(\phi;k)-(N-1)]\nonumber\\
          & &+V_2(\phi,G_{\sigma},G_{\pi})~,
\end{eqnarray}
where the last term originates from the  double bubble diagrams 
and its  contribution reads as
\begin{equation}
V_2(\phi,G_{\sigma},G_{\pi})=
3\frac{\lambda}{6N}\left [
\int_{\beta}G_{\sigma}(k)\right ]^{2}
+\frac{\lambda(N^{2}-1)}{6N}\left [
\int_{\beta} G_{\pi}(\phi;k)\right ]^{2}
+\frac{\lambda(N-1)}{6N}\left [\int_{\beta} G_{\sigma}(\phi;k)\right ]
\left [\int_{\beta}G_{\pi}(\phi;k)\right ]~.
\end{equation}
The weight factors appearing in the above expression  can be understood
in a similar way as in the $O(4)$ case  the only difference being
the $N-1$ pion fields. Of course it is easy to see that we recover the 
previous case by simply substituting $N=4$.

As in the case of $\lambda\phi^{4}$ and the $O(4)$ model 
we minimise  the effective
potential with respect to the dressed propagators and we get 
a set of gap 
equations. By using the same form  
for the dressed propagators as before, we end 
up with the following set of nonlinear gap
equations for the thermal effective particle masses
\begin{eqnarray}
&&M_{\sigma}^{2}=m^{2}+\frac{2\lambda}{N}\phi^{2}+
\frac{2\lambda}{N}F_{\beta}(M_{\sigma})
+\frac{2\lambda(N-1)}{3N}F_{\beta}(M_{\pi})\nonumber\\
&&M_{\pi}^{2}=m^{2}+\frac{2\lambda}{3 N}\phi^{2}
+\frac{2\lambda}{3N}F_{\beta}(M_{\sigma})
+\frac{2\lambda(N+1)}{3N}F_{\beta}(M_{\pi})~,
\label{masseslargen}
\end{eqnarray}
where we only keep the finite temperature part of 
the integrals. As it is easy to observe, for $N=4$ we obtain 
identical expressions for the system of gap  equations 
as in the case of the $O(4)$ model. Then the effective potential
will appear in the form
\begin{eqnarray}
V(\phi,M)&=&\frac{1}{2}m^{2}\phi^{2}+\frac{1}{6N}\lambda\phi^{4}
+ \frac{1}{2}\int_{\beta}\ln(k^{2}+M_{\sigma}^{2})
+ \frac{(N-1)}{2}\int_{\beta}\ln(k^{2}+M_{\pi}^{2})
\nonumber\\
         &&\mbox{}- \frac{1}{2}(M_{\sigma}^{2}-m^{2}
           -\frac{2\lambda}{N}\phi^{2})F_{\beta}(M_{\sigma})
   -\frac{N-1}{2}(M_{\pi}^{2}-m^{2}-\frac{2\lambda}{3 N}\phi^{2})
F_{\beta}(M_{\pi})\nonumber\\
 &&\mbox{}+\frac{\lambda}{2N}[F_{\beta}(M_{\sigma})]^{2}
+\frac{\lambda(N^{2}-1)}{6N}[F_{\beta}(M_{\pi})]^{2}
+\frac{\lambda(N-1)}{3N}F_{\beta}(M_{\sigma})F_{\beta}(M_{\pi})~.
\end{eqnarray}

In the large $N$ approximation, which means that we ignore
terms of $O(1/N)$, the system of 
the two equations (\ref{masseslargen}) reduces to
\begin{eqnarray}
&&M_{\sigma}^{2}=m^{2}+\frac{2\lambda}{N}\phi^{2}
+\frac{2\lambda}{3}F_{\beta}(M_{\pi})
\nonumber\\
&&M_{\pi}^{2}=m^{2}+\frac{2\lambda}{3 N}\phi^{2}
+\frac{2\lambda}{3}F_{\beta}(M_{\pi})~.
\end{eqnarray}
We have retained the terms  quadratic in $\phi$ 
since $\phi$ depends on $N$ as $\phi^{2}=3 N m^{2}/2\lambda$
and so these terms are of $O(1)$. In order to solve this
system and be in ``some contact'' with phenomenology in the chiral
limit, we set now 
$N=4$ so at zero temperature the pions are massles and sigma 
has a mass $M^{2}_{\sigma}=-2m^{2}$. Now our system is written as
\begin{eqnarray}
&&M_{\sigma}^{2}=m^{2}+\frac{1}{2}\lambda\phi^{2}
+\frac{2\lambda}{3}F_{\beta}(M_{\pi})
\nonumber\\
&&M_{\pi}^{2}=m^{2}+\frac{1}{6}\lambda\phi^{2}
+\frac{2\lambda}{3}F_{\beta}(M_{\pi})~,
\label{mp}
\end{eqnarray}
and in order to solve it we proceed in a similar way as in Hartree
approximation.

At very high temperatures the potential has only one
minimum that at $\phi=0$ and in this case the two equations
become degenerate
\begin{equation}
M_{\sigma}^{2}=M_{\pi}^{2}= M^{2}=m^{2}+\frac{2\lambda}{3}F_{\beta}(M)~.
\end{equation}
This last equation actually defines the critical 
temperature. $F_{\beta}(M)$ is given by the same expression
as in the $O(4)$ case. The mass of the particles vanishes 
at the critical temperature, so we can use the result of eqn. (\ref{FBM})
to find that the critical temperature is at
\begin{equation}
T_{c}=\sqrt{3}\left (-\frac{6 m^{2}}{\lambda} \right )^{1/2}
=\sqrt{3}f_{\pi}\approx 161 \rm MeV~. 
\end{equation}

Before proceeding to examine the low temperature phase we should 
note an observation which actually
marks the significant difference between the Hartree approximation 
in the $N=4$ case and the large $N$ approximation. Minimizing  
the potential with respect 
to $\phi$ gives
\begin{equation}
{dV(\phi,M)\over d\phi}={\partial V \over \partial \phi}=\phi\left[m^{2}
+\frac{2\lambda}{3 N}\phi^{2}+\frac{2\lambda}{3N}F_{\beta}(M_{\sigma})
+\frac{2\lambda(N-1)}{3N}F_{\beta}(M_{\pi})\right]=0~,
\end{equation}
which in the large $N$ approximation becomes
\begin{equation}
{dV(\phi,M)\over d\phi}=\phi\left[m^{2}+\frac{\lambda}{6}\phi^{2}
+\frac{2\lambda}{3}F_{\beta}(M_{\pi})\right]=0~.
\end{equation}
Combining this last equation with the second of
equations (\ref{mp}) above we observe that
\begin{equation}
{dV(\phi,M)\over d\phi}=\phi M^{2}_{\pi}=0~.
\end{equation}
Therefore  the large $N$ approximation  
implies that the pions should be massless in the low 
temperature phase in accordance
with the Goldstone theorem. 

This observation is reflected in the solution of the system 
of the gap equations as it is shown in fig. 6a. The pions at 
low temperatures appear 
as massless, but at high temperatures the thermal contribution 
to the effective masses
make them degenerate with the sigma. The order
parameter vanishes continuously in this case as it is shown in fig. 6b
and this corresponds to a second order phase transition.
\begin{figure}[h]
\begin{center}
\mbox{
\epsfxsize=10cm
\epsfbox{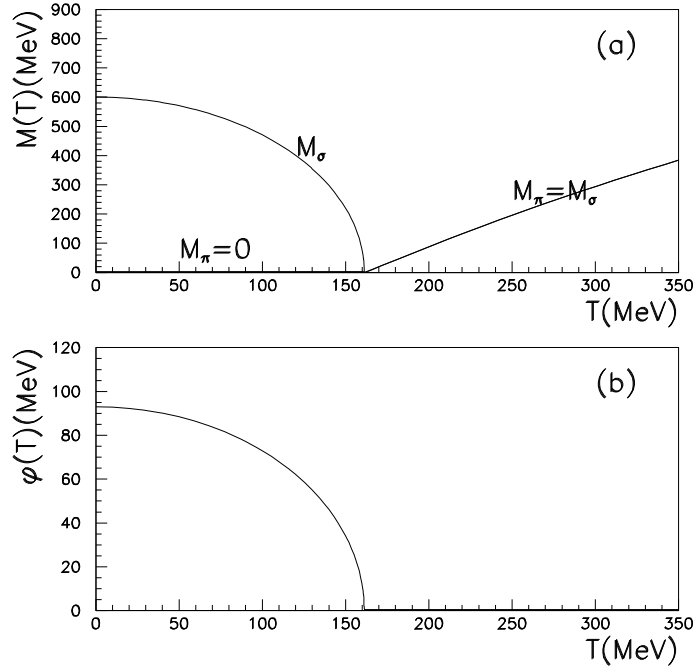}
}
\end{center}
\caption{(a) Solution of the system of gap equations in the large $N$
approximation in the  chiral limit. At low temperatures
the pions appear as massless. (b) Evolution of the order parameter with
temperature in the large $N$ approximation.}
\end{figure}
 
\subsection{Large $N$ approximation in the broken 
symmetry case $\varepsilon \neq 0$}

As already mentioned  for the $O(4)$ case, the 
symmetry breaking 
term $\varepsilon\sigma$ has been 
introduced into the Lagrangian in order to generate the 
observed masses of the pions. The same can be done
for the $O(N)$ model the only difference being the
$N-1$ pion fields. Inserting this term into the expression 
for the effective potential and 
differentiating with respect to $\phi$ we 
obtain, as in the $O(4)$ case, one more equation. In
the large $N$ approximation this is written as
\begin{equation}
\left [ m^{2}+\frac{1}{6}\lambda\phi^{2}
+\frac{2\lambda}{3}(F_{\beta}(M_{\pi}) \right ]\phi-\varepsilon=0~.
\label{largenf}
\end{equation}

We have solved this last system of equations (\ref{mp}) and (\ref{largenf})
numerically and the solution is given in fig. 7. As in the $N=4$ case 
there is no longer any phase transition. We encounter again the 
crossover phenomenon between
the low and high temperature phases, the difference being now
that the change of the order parameter (fig. 7b) in the transition 
region is much more smooth  in contrast the 
more ``sharp'' behaviour seen  in the 
$N=4$ case in fig. 5b. 
 
\begin{figure}[h]
\label{nbreak}
\begin{center}
\mbox{
\epsfxsize=10cm
\epsfbox{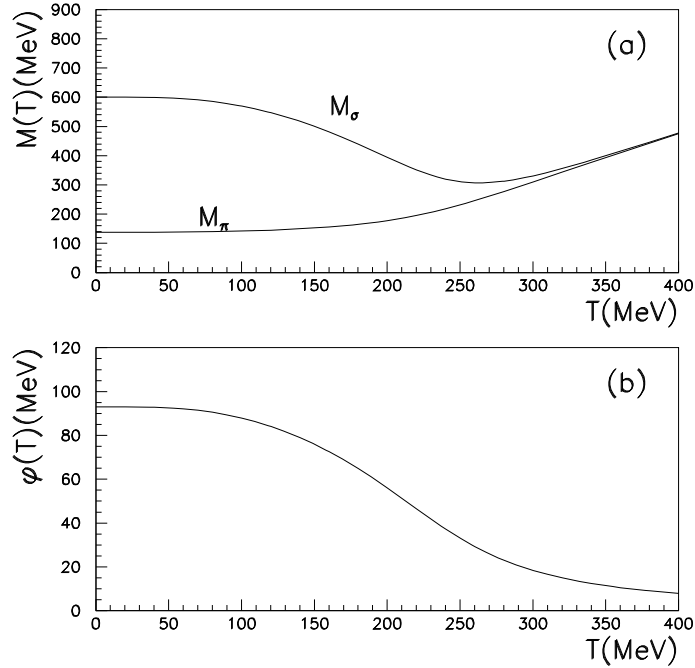}
}
\end{center}
\caption{(a) Solution of the system of gap equations in the large $N$
approximation in the case of broken chiral symmetry
$\varepsilon \neq 0$. At low temperatures the pions appear with
the observed masses. (b) Evolution of the order parameter with
temperature.}
\end{figure}

\section{Conclusions}

We have studied 
the chiral phase transition
using the linear sigma model. In order to get some insight  
into how the phase transition could proceed, we have 
calculated the finite temperature effective
potential of this model in the Hartree and large $N$ 
approximations using the CJT formalism of composite 
operators. The method proved to be very handy since we 
actually need 
to calculate only one diagram. In both cases we 
have solved numerically the system
of gap equations and found  the evolution with temperature of 
the effective thermal masses. In the Hartree 
approximation  we find a first order phase 
transition, but in contrast to that, the large $N$
approximation predicts a second order phase transition. This
last observation seems to be in agreement with
different approaches to the chiral phase transition based 
on the argument that the linear sigma model belongs
in the same  universality class as other models which
are known  to exhibit second order phase 
transitions \cite{rajagopal,rw399}. However in the large
$N$ approximation the sigma contribution is ignored and
this of course introduces errors when we
calculate  the critical temperature. In the case $N=4$
which is closer to phenomenology we could probably obtain 
a better approximation if we had
considered the effects of interactions given by the last two terms
in the Lagrangian (\ref{lint}). We are planning
some investigations in order  to include the effects of these
terms in the calculation of the effective potential.	 
 
When we include the symmetry breaking term $\varepsilon \sigma$
which generates the pion observed masses, both in Hartree and 
large $N$ approximations we found that there is no longer any phase
transition. We rather observe a crossover phenomenon where the 
change of the order parameter in the Hartree case occurs more rapidly
in contrast to the more smooth behaviour  exhibited by the large
$N$ approximation.  Our observation confirms
results reported recently by Chiku and Hatsuda \cite{chiku}
using a different approach. In their analysis they also report
indication of first order phase transition in the chiral limit.

Of course as we have already pointed out, the linear 
sigma model is only an approximation
to the real problem which is QCD, but the study of the 
chiral phase transition in the framework of this model
could be a helpful guide to how one could tackle the 
original problem and get some insight in the physics 
invloved. We have used the imaginary time formalism
which is adequate for studies at thermal equilibrium 
but if one is interested in studies of the dynamics of the phase 
transition, the real time formalism seems to be more 
convenient \cite{smilga,lebellac,landsman}. We are currently making
some preliminary investigations in this direction.

\acknowledgements{The author would like to express his gratitude to
Mike Birse for his continuous guidance at all stages of
this work and Joannis Papavassiliou for helpful conversations
at the early stages of this work. Support from  EPSRC is 
also acknowledged.}

\end{document}